\documentclass[%
 aip,
 amsmath,amssymb,
reprint,%
floatfix,
]{revtex4-1}

\usepackage{graphicx}
\usepackage{dcolumn}
\usepackage{bm}

\usepackage[utf8]{inputenc}
\usepackage[T1]{fontenc}
\usepackage{mathptmx}
\usepackage{etoolbox}
\usepackage{blindtext}

\makeatletter
\def\@email#1#2{%
 \endgroup
 \patchcmd{\titleblock@produce}
  {\frontmatter@RRAPformat}
  {\frontmatter@RRAPformat{\produce@RRAP{*#1\href{mailto:#2}{#2}}}\frontmatter@RRAPformat}
  {}{}
}%
\makeatother
\begin{document}

\preprint{AIP/123-QED}

\title[]{\textit{In-situ} High Pressure Nuclear Magnetic Resonance Crystallography}

\author{Thomas Meier}
   \email{thomas.meier@hpstar.ac.cn}
   \affiliation{Center for High Pressure Science and Technology Advance Research, Beijing, China}
   
\author{Alena Aslandukova}
  \affiliation{Bayerisches Geoinstitut, University of Bayreuth, Bayreuth, Germany}

\author{Florian Trybel}
  \affiliation{ Department of Physics, Chemistry and Biology (IFM), Link{\"o}ping University, SE-581 83, Link{\"o}ping, Sweden} 

\author{Dominique Laniel}
  \affiliation{Material Physics and Technology at Extreme Conditions, Laboratory of Crystallography, University of Bayreuth,  Bayreuth, Germany}

\author{Takayuki Ishii}
  \affiliation{Center for High Pressure Science and Technology Advance Research, Beijing, China}

\author{Saiana Khandarkhaeva}
  \affiliation{Bayerisches Geoinstitut, University of Bayreuth,  Bayreuth, Germany}
   
\author{Natalia Dubrovinskaia}
  \affiliation{Material Physics and Technology at Extreme Conditions, Laboratory of Crystallography, University of Bayreuth,  Bayreuth, Germany}                         

\author{Leonid Dubrovinsky}
  \affiliation{Bayerisches Geoinstitut, University of Bayreuth, Bayreuth, Germany}  

\date{\today}

\begin{abstract}
Our recent developments in \textit{in-situ} nuclear magnetic resonance (NMR) spectroscopy under extreme conditions led to the observations of a wide variety of physical phenomena not accessible with standard high pressure experimental probes. However, inherent di- or quadrupolar line broadening in diamond anvil cell (DAC) based NMR experiments often limit detailed investigations of local atomic structures, especially if different phases or local environments are coexisting. Here, we present our progress in the development of high resolution NMR experiments in DACs using one- and two-dimensional homonuclear decoupling experiments at pressures up to the Mbar regime. Using this technique, spectral resolutions in the order of 1 ppm and below have been achieved, enabling high pressure structural analysis. Several examples will demonstrate the wide applicability of this method for extreme conditions research.
\end{abstract}

\maketitle

\section{Introduction}
Nuclear magnetic resonance (NMR) spectroscopy is widely considered one of the most versatile spectroscopic methods available to contemporary natural sciences \cite{Levitt2000}. Its ability to detect small variations in the nuclear Larmor precession frequency of a $I>0$ nucleus makes NMR measurements an invaluable tool for electronic and atomic structure determination\cite{Grant2007}.
\\
Contrary to X-ray based crystallography, probing long range structural symmetries in solids, quasi-solids and powders\cite{Dubrovinskaia2018}, NMR crystallography is a well established method identifying, and often even quantifying, local short range atomic surroundings or sub-units in macro-molecules, like proteins\cite{MacArthur1994, Bryce2017, Martineau2014}.
\\
Applying NMR towards extreme conditions research\cite{Mao2016}, in particular at tens of GPa, was widely considered impossible due to the severe technical difficulties in implementing NMR in high pressure generating devices, on the one hand, and, on the other hand, the inherently low spin sensitivities of the polarised nuclear spin ensemble in an external magnetic field\cite{Meier2017b}.\\
Our recent developments using magnetic flux focusing Lenz lenses showed that NMR experiments are possible at pressures up to several hundred of GPa \cite{Meier2017,Meier2018a, Meier2019a} and temperatures above 1000 K\cite{Meier2019b}. 
This method, despite enabling reaching significantly higher pressures than previous NMR configurations, is generally prone to low spectral resolution limiting its applicability to detect pronounced electronic or magnetic transitions\cite{Meier2018b}. Determination of different crystallographic structures, coordination sites or local environments also remains challenging.\\
Here we present our recent advancements leading to the development of high resolution one and two dimensional NMR techniques in DACs. This method was shown to provide intricate structural and electronic insights for a wealth of different systems, namely diatomic molecules ($^{1}H$-NMR on $H_2$ and $^{14}N$-NMR on $N_2$), pressure- and temperature-formed metal hydrides ($^{1}H$-NMR on $YH_x$ systems) as well as on geophysically important materials ($^{1}H$-NMR on $(Al_{0.3}, Fe_{0.7})OOH$ and dense magnesium silicate phase D).   
\section{Static Resonance Line Narrowing Techniques}
\begin{figure}[htb]
\includegraphics[width=0.45\textwidth]{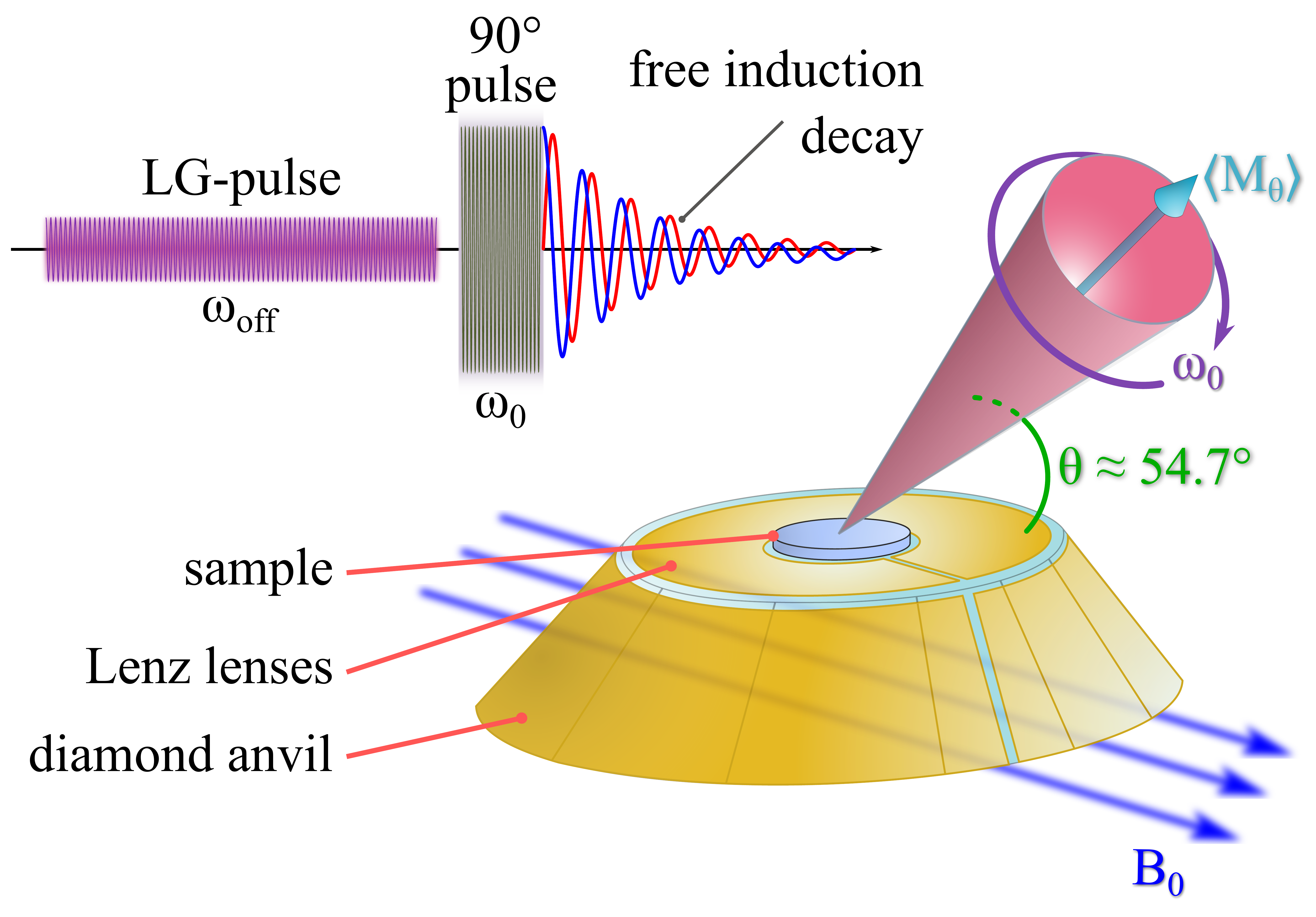}%
\caption{\textbf{Schematic representation of Lee-Goldburg decoupling pulse experiments in DACs.} By irradiating the sample with a long and weak off-resonant pulse, the precessing spin system is forced to relax in the magic angle $\Theta \approx 54.7 ^{\circ}$ (see text for details), effectively averaging out dominant non-secular spin interactions. The resulting free induction decay in the rotating frame (FIDRF) will only be subject to spin interactions linear in the nuclear Zeeman interaction perturbation (\textit{e.g.} isotropic chemical, paramagnetic or Knight shifts).}   
\label{fig:schematic}
\end{figure} 
The effects leading to significant line-broadening in DAC-based NMR research can be divided in two categories: 1) external line broadening effects such as partial shielding or distortion of the external polarizing magnetic field $B_0$ by the use of paramagnetic DAC components and 2) nuclear spin interactions inherent to a specific compound. While the former effects can easily be minimized by the use of different materials, the suppression of the latter prove to be difficult.
\\
The most dominant nuclear spin interactions leading to resonance frequency dispersion are interactions due to homo- or heteronuclear dipole-dipole interactions (${\cal H} _{DD}$), coupling of the quadrupole moment of $I>1/2$ nuclei with the surrounding electric field gradient of the charge symmetry (${\cal H}_Q$), interaction with paramagnetic centers (${\cal H}_{PM}$) as well as anisotropies in dia- and paramagnetic shielding tensors (${\cal H}_{CSA,K}$).\\
To a first order approximation, analytic expressions of these interactions can be rearranged in a form containing orientation and distance dependencies as well as a contribution consisting of information about their respective magnitude $\Xi(I)$\cite{Smith1992}:
\begin{equation}
    {\cal H}_i (I,r_{IJ},\theta) = \frac{3\cos^2(\theta) - 1}{2r_{IJ}^3}\cdot \Xi(\textit{I}),
     \label{eq1}
\end{equation}
The angle $\theta$ in equation (\ref{eq1}) describes the orientation of the nuclear magnetic dipoles relative to the external magnetic field. Powder averaging over this term leads to characteristic line shapes well known in NMR spectroscopy\cite{Pake1948}.\\ 
Noticeably, all of these spin interactions show a pronounced inverse cubic dependence on atomic distances, $r_{IJ}$. Therefore, the application of pressure often leads to drastically increased interaction energies, as atomic separations are diminished and thus significantly increased resonance line-widths.\\
The angular dependence of equation (\ref{eq1}) has a zero crossing at an angle of $54.7^{\circ}$. Thus, an alignment of all nuclear dipoles towards this so-called "magic angle" would effectively suppress the most dominant line broadening effects in solids. Two strategies have been developed to accomplish this feat.\\
Mechanical fast rotation of a sample around the magic angle, so-called magic angle spinning (MAS), is nowadays in specialised NMR probes the most widespread method for resonance line narrowing, allowing for spectral resolutions in the order of 0.1 ppm\cite{Hennel2005}. For efficient line narrowing in MAS, the rotation speed of the sample around the magic angle should be much greater than the dominant spin interaction frequencies. For high pressure experiments, line-widths in the order of several hundred kHz have repeatedly been reported\cite{Meier2014,Meier2018, Meier2020}; far outside of the capabilities of commercially available MAS probes. \\
In an alternative approach, first introduced by Lee and Goldburg\cite{Lee1965}, the spin system itself is allowed to precess along the magic angle by the application of a low amplitude off-resonance pulse (\textit{i.e.} the LG pulse, see figure \ref{fig:schematic}). Signal acquisition is achieved by sampling the zero-time free induction decay intensities in the laboratory frame for different time increments of the Lee-Goldburg (LG) pulse. The resulting free induction decay in the rotating frame (FIDRF) will be solely dependent on spin interactions linear in the nuclear Zeeman interaction, such as isotropic chemical, paramagnetic or Knight shifts.
\\
While this approach has the advantage that it can be conducted using NMR equipment without rapid mechanical rotation (\textit{i.e.} static), the necessary radio frequency field amplitudes and bandwidths of standard probes often limits its effectiveness. However, high pressure NMR set-ups are usually equipped with rather small resonators--and thus large radio frequency field amplitudes--capable of sustaining large bandwidths\cite{Meier2015}.
\\
In a Lee-Goldburg NMR experiment, an effective radio frequency magnetic field $\vec{B}_{eff}$ will be generated by the NMR coil:
\begin{equation}
    \vec{B}_{eff}=\left( B_0-\frac{\omega_{off}}{\gamma_n}\right)\cdot \vec{k} + B_1 \cdot \vec{j}
    \label{eq2}
\end{equation}
In equation (\ref{eq2}) $B_0$ is the external magnetic field amplitude, $\omega_{off}$ is the off-resonance frequency of the LG pulse, $\gamma_n$ is the gyromagnetic ratio of the NMR probe nucleus and $B_1$ is the radio frequency field amplitude of the NMR coil. $\vec{k}$ and $\vec{j}$ are unit vectors along the direction of $B_0$ and $B_1$ respectively. The most general expression for $B_1$ generated by solenoidal coils is: 
\begin{equation}
    B_1=\sqrt{\frac{\mu_0 Q P_{pulse}}{2\omega_0 V_{coil}}}
\end{equation}
where $\mu_0$ is the vacuum permeability, Q is the resonance circuits quality factor with $Q=\omega_0 L/R$ ($\omega_0$ is the nuclear Larmor frequency, $L$ is the inductance and $R$ the AC resistance of the tank circuit), $P_{pulse}$ is the average pulse power and $V_{coil}$ can be approximated to be the inner volume of the coil.\\
The angular dependence of equation (\ref{eq1}) is zero for $\theta=\arccos(1/\sqrt{3})$, thus the angle between $\vec{B}_{eff}$ and $\vec{B}_{0}$, using (\ref{eq2}), becomes:
\begin{equation}
    \tan\theta = \sqrt{2} = \frac{B_1}{B_0-\frac{\omega_off}{\gamma_n}}
\end{equation}
Therefore, the frequency offset ($\Delta \omega$) in rad/s of the LG pulse relative to the spectrometer frequency can be obtained -- using the resonance condition for NMR $\omega_0=\gamma_n B_0$ -- via:
\begin{equation}
    \Delta \omega=\omega_0-\omega_{off}=\frac{\gamma_n B_1}{\sqrt{2}}
    \label{woff}
\end{equation}
For \textit{in-situ} NMR in DACs using Lenz lens based radio-frequency spin excitation, the $B_1$ amplitude of the driving Helmholtz coil arrangements is typically in the order of 1 mT, leading to offset frequencies of about 30 kHz at $B_0 = 7.04~T$ which is well within the capabilities of high pressure NMR probes. Therefore, Lee-Goldburg based homonuclear spin decoupling experiments appear as a natural, and very powerful, tool for high pressure NMR experiments where high spectral resolutions are necessary\cite{Meier2019}. 
\section{Experimental}
\begin{figure*}
\includegraphics[width=0.65\textwidth]{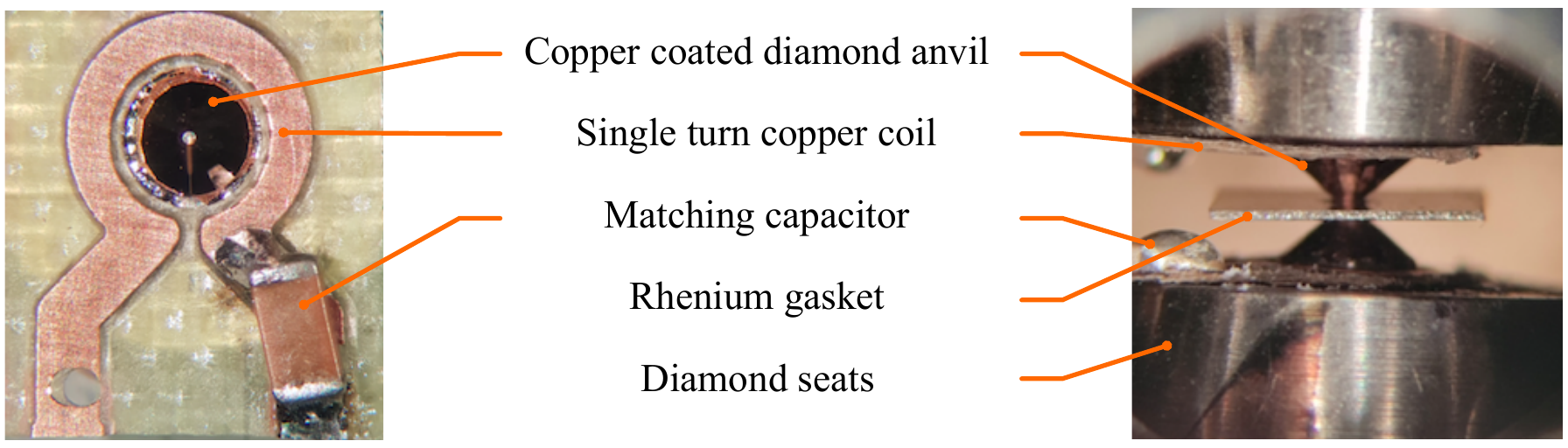}%
\caption{\textbf{High Pressure NMR resonator set-up for high frequency applications.} \textbf{Left}: To drive the Lenz lens resonators, a pair of single loop coils made from printed circuit board (PCB) plated copper is used. In order to match the resonators' small inductance ($\approx 1~nH$) to the desired resonance, a SMD (surface-mount device) capacitor ($\approx 10~nF$) was placed in series with each coil. \textbf{Right}: In the closed DAC assembly, a Helmholtz coil arrangement drives the Lenz lens resonators.}
\label{fig:resonator}
\end{figure*}
The DACs for high pressure NMR experiments were prepared following a procedure close to the typical one. First, rhenium gaskets were indented to the desired thickness, which depends on the size of the diamond anvil culets employed, but usually not more than 25 $\mu m$. Sample cavities were drilled using specialized laser drilling equipment. After gasket preparation, the diamond anvils were covered with a layer of 1 $\mu m$ of copper or gold using chemical vapour deposition. To ensure the electrical insulation of the conductive layers from the rhenium gasket, the latter were coated by a thin layer ($\approx 500~nm$) of $Al_2O_3$ using physical vapour deposition. The Lenz lens resonators were shaped from the conductive layer on the diamonds by using focused ion beam milling. 
\\
Before the final cell assembly, radio frequency resonators were prepared accordingly to their desired operation frequency. A pair of high inductance solenoid coils ($\approx 100~nH$) for low frequency experiments of below 100 MHz or a pair of single turn printed circuit board (PCB) plated copper resonators for $^{1}$H-NMR frequencies at high fields (\textit{e.g.} 300 MHz) were used as driving coil arrangements for the Lenz lens resonators' structure and were placed around each diamond anvil, as seen in figure \ref{fig:resonator}. After sample loading and initial pressurisation, the driving coils were connected to form a Helmholtz coil-like arrangement.
\\
Pressure calibration was performed using the shift of the first derivative of the first order Raman signal of the diamond edge in the center of the culet \cite{Akahama2004, Akahama2006}. All DACs were fixed and connected to home built NMR probes equipped with customized cylindrical trimmer capacitors (dynamic range of $\approx~150$ pF) for frequency tuning to the desired resonance frequencies and impedance matching to the spectrometer electronics ($50~\Omega$). All experiments were conducted at a magnetic field of 7.04 T, corresponding to $^1$H frequencies of 300 MHz.
\\
Proton shift referencing was conducted by using the $^{63}$Cu resonances of the Lenz lenses themselves as internal references taking into account the additional shielding of $B_0$ inherent to every DAC. These resonances were cross referenced with standard metallic copper samples at ambient conditions without a DAC. The resulting shift between both $^{63}$Cu-NMR signals are then used as a primer for the NMR signals of the samples under investigation.
\\
Lee-Goldburg decoupling experiments were initially calibrated by quick two-dimensional nutations ($\sim$ 56 2D spectra) for different off resonant frequencies of the LG-pulse. Optimal $\omega_{off}$ values were found between 25 to 35 kHz, in agreement with our calculations (eq. (\ref{woff})). One dimensional LG-spectra were recorded by oversampling in the indirect time domain ($\sim$ 8000 increments) using previously determined values of $\omega_{off}$ at a 10 dB pulse power attenuation relative to the excitation pulse. Two dimensional LG spectra were recorded with identical direct and indirect time domains (usually 2048 points in each dimension), while matching the incrementation of the LG-pulse to the direct time domain dwell time of the spectrometer. This procedure ensures correct scaling of the indirect LG frequency domain after 2D-Fourier transform\cite{Ernst1989}.
\begin{figure*}
\includegraphics[width=0.75\textwidth]{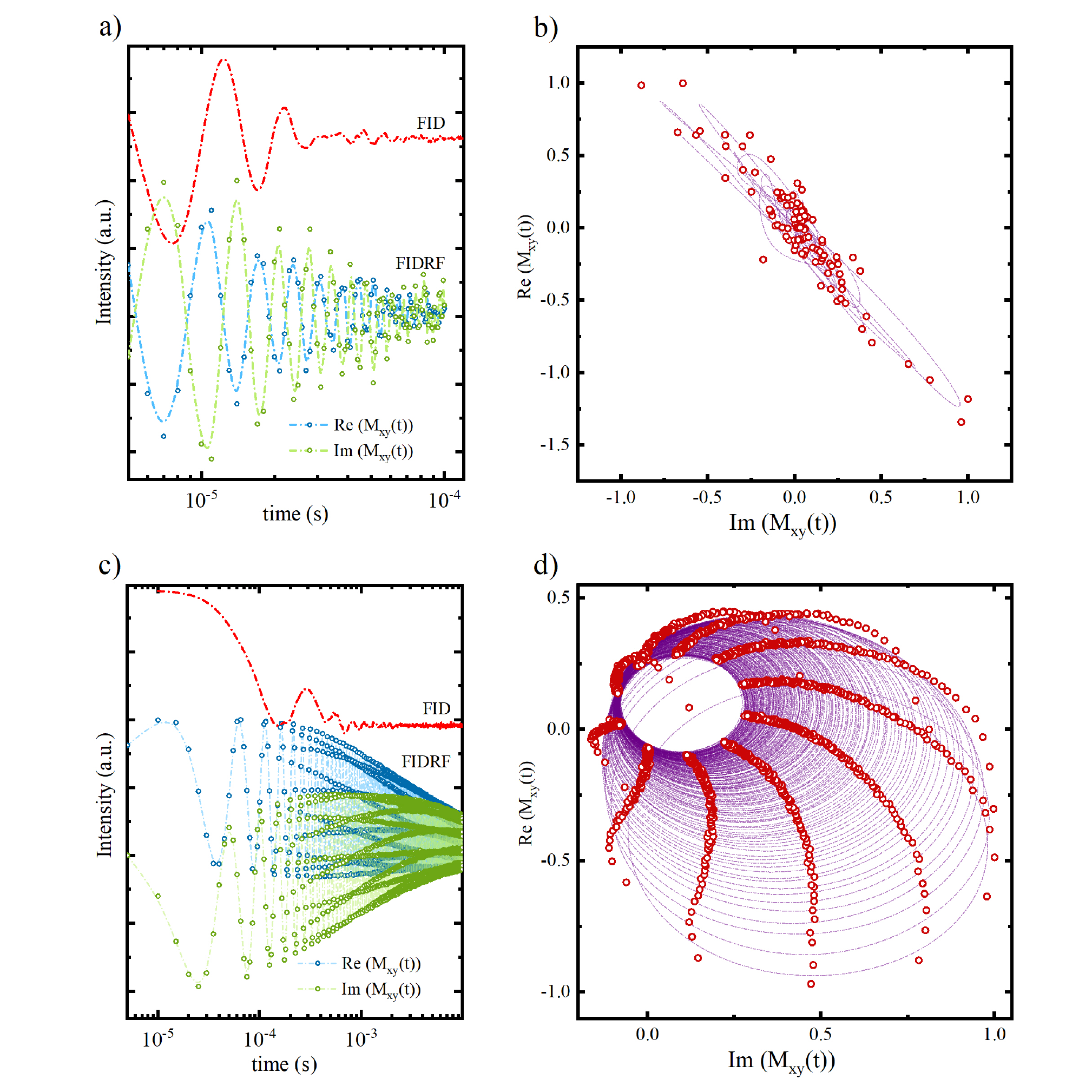}%
\caption{\textbf{Comparison of LG-decoupling using standard coils and in DACs.} \textbf{a}) $^{19}$F-free induction decays in the laboratory frame (FID) and in the rotating frame (FIDRF), sampled as described in the text, of a single crystal of $CaF_2$ obtained from a 150 pl (500~$\mu m$ diameter, 750~$\mu m$ height) sized solenoidal coil.
\textbf{b}) The corresponding coherence diagram (red dots are data points, the purple lines are guide to the eye) clearly shows a rapid dephasing of the FIDRF within $\sim100\mu s$ \textbf{c}) $^{1}$H FID and FIDRF of a single crystal of dense magnesium silicate phase D from a Lenz lens based resonator design as described in the text. d) The corresponding coherence diagram shows a slow dephasing of real and imaginary parts of the FIDRF, corresponding to a streching factor of about two orders of magnitude. The resulting Fourier transform NMR signal (not shown) has a FWHM line width of 0.12 ppm.} 
\label{fig:LGcompare}
\end{figure*}

\section{ \textit{Performance of LG decoupling with Lenz lens based NMR resonators}}

In order to demonstrate the advantages of LG decoupling in DAC-based NMR experiments, we compare the LG-performance between a standard solenoidal microcoil ($500~\mu$m diameter, $750~\mu$m height) and Lenz lens based NMR resonators. Ideal LG-conditions were determined in both cases.
\\
First, we performed one-dimensional $^{19}$F-LG-NMR experiments on a single crystal of CaF$_2$ by incrementing the saturation pulse in steps of 1 $\mu$s (maximum length of the LG-pulse was 100 $\mu$s) at $\Delta \omega\simeq 100$ kHz.  The zero-time intensity sampling of both real and imaginary parts of the signals (Figure \ref{fig:LGcompare}a) decay within approximately 125 $\mu$s, corresponding to a 5-fold stretching of the FIDRF relative to the laboratory frame (FID). The resulting line widths of the fluorine signals after Fourier transform could be reduced from initally 140 ppm to about 30 ppm. The correlation between real and imaginary parts of the FIDRF (\ref{fig:LGcompare}b) shows the rapid de-phasing of the NMR signal in the rotating frame.
\\
Subsequently, we performed the same experiment on a single crystal of dense magnesium silicate phase D in a DAC at 9 GPa, using LG-pulse increments of 5 $\mu$s at $\Delta \omega \simeq 34$ kHz. The resulting FIDRF (Figure \ref{fig:LGcompare}c) exhibits a very slow decay of signal amplitudes. Data acquisition in the indirect LG time domain was cut off before complete signal decay was achieved due to spectrometer limitations. Analysis of the FIDRF indicates a $1/e$ decay constant of about 40 ms. The corresponding coherence diagram (Figure \ref{fig:LGcompare}d) shows ideal vortex-like behaviour of slowly de-phasing signal amplitudes. FWHM line widths of the resulting Fourier transform $^{1}H$-NMR signals were found to be 0.12 ppm. 
\\
This comparison demonstrates that homonuclear decoupling in DACs benefits from strong radio frequency field amplitudes $B_1$ and large excitation and receiving bandwidths of the employed resonator configuration. It is noteworthy that the micro-coil used for experiments on CaF$_2$ is about four orders of magnitude smaller than regular NMR transceiver coil arrangements which would lead to even lower $B_1$ amplitudes and bandwidths.
\\
\begin{figure*}[htb]
\includegraphics[width=0.65\textwidth]{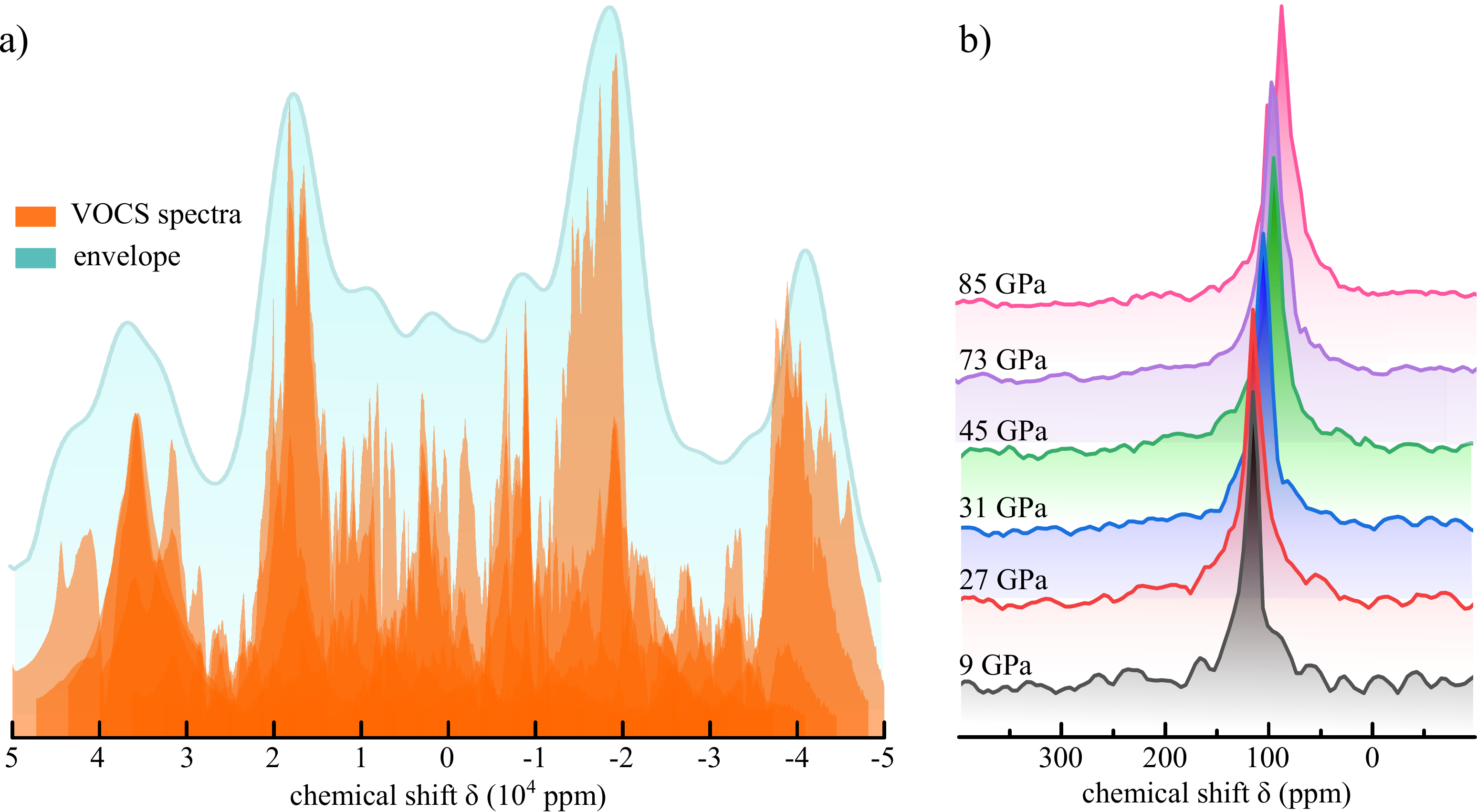}%
\caption{\textbf{$^{14}$N-LG-NMR spectra of molecular nitrogen up to 85 GPa}.
\textbf{a}) The quintuplet state of molecular Nitrogen, $^{14}N_2$, possess a nuclear molecular spin of I=2, leading to significantly broadened spectra of about 2 MHz ($\approx 10^{5}$ ppm). The full spectrum is a sum of spin echoes (orange) acquired at variable frequency offsets. The blue spectrum is a broadened envelope of all sub-spectra.
\textbf{b}) By the application of LG-decoupling it was possible to resolve isotropic chemical shifts of molecular nitrogen at ultra high densities with an accuracy of $\sim$ 10 ppm.
\label{fig:nitrogen}
}
\end{figure*} 
\section{Results}
In the following, we give examples of measurements where LG decoupling enables the investigation of:
\begin{description}
\item[(i)] spin isomerism in molecular hydrogen and nitrogen, which can only be resolved in a one dimensional analysis due the strong decrease in line width.
\item[(ii)] a subtle spin transition in Phase-D, via a two dimensional analysis, which is not resolved in 1D or by other experimental techniques.
\item[(iii)] coexisting local chemical environments in $\delta$-(Al,Fe)OOH and different coexisting phases, e.g. YH$_2$ and YH$_3$, in one experiment where a two dimensional analysis leads to a decomposition of the resonances and enables therefore not only an individual analysis, but also a simultaneous analysis at the same $P$ and $T$, as well as, a quantification of the sample fractions.
\end{description}
\subsection{One dimension}
\subsubsection{$^{1}$H-LG-NMR on molecular hydrogen up to 123 GPa}
First LG-decoupling experiments in DACs have been published recently on molecular hydrogen with resulting spectral resolutions of about 3 ppm corresponding to a 1600-fold reduction of the H$_2$ NMR signals at pressures of up to 123 GPa \cite{Meier2020}.
\\
Molecular hydrogen is predicted to show a coupling between molecular rotational properties and nuclear spin orientations, giving rise to the spin isomers ortho- and para-hydrogen. We were able to show that even at already at $P \gtrsim 70$ GPa intramolecular nuclear spin coupling breaks down and the hydrogen spin system adopts an average dipolar $I=1/2$ value, much lower than the predicted molecular dissociation at the Wigner-Huntington transition\cite{Gregoryanz2020, Dias2016a, Geng2017}. Crossovers of the nuclear spin statistics of a quantum solids such as hydrogen have never been observed before and given the large compressibility of hydrogen in conjunction with strong nuclear quantum effects, this crossover phenomenon might only be experimentally observable in molecular H$_2$.
\\
\subsubsection{\textit{$^{14}$N-LG-NMR on molecular nitrogen up to 85 GPa}}
Similar to molecular diatomic hydrogen, molecular $N_2$\cite{Laniel2020, Li2018} and its derived nitrides\cite{Laniel2019, Laniel2018a, Laniel2018}
attracted a great interest in the high pressure community as high energy density materials. Nitrogen-NMR experiments at extreme conditions would be greatly beneficial for investigations of the local electronic and atomic environments of these compounds, but pronounced quadrupolar couplings, low signal intensities, structural complexity and the necessity for laser heating assisted sample synthesis in a DAC under pressure makes Nitrogen-NMR in DACs extremely challenging.    
\\
At natural abundances, the majority of nitrogen molecules will be $^{14}$N$_2$ units, with each $^{14}$N nucleus possessing integer nuclear spin ($I=1$). Intramolecular nuclear angular momentum summation leads to the stabilisation of two observable nuclear spin isomers: (i) the quintuplet state of spin $I=2$ ( $|I=2,~m_I=\pm 2,\pm 1,0\rangle$) and  (ii) the $I=1$ spin triplett ($|I=1,~m_{\rm I}=\pm 1,0\rangle$), where the singlett state of $I=0$ is NMR inactive and not observable. The quadrupolar nature of these spin isomers leads to significantly broadened $^{14}$N-NMR spectra of obout $\geq 10^{5}$ ppm \cite{ODell2010}. As NMR signal intensities are proportional to the number density of each spin species within the sample, high pressure NMR experiments on molecular nitrogen are extremely challenging considering usual sample volumes in the order of 10 pl.
\\
First preliminary $^{14}$N-NMR experiments on molecular nitrogen at 72 GPa after laser heating to about 2000 K at 130 GPa displayed a 2 MHz broad ($\approx 10^{5}$ ppm) $^{14}$N-NMR spectrum, see figure \ref{fig:nitrogen}a. The full spectrum was acquired using the well established variable offset cumulative spectroscopy (VOCS) method\cite{Schurko2013}.
\\
Figure \ref{fig:nitrogen}b shows high resolution spectra between 9 and 85 GPa. Considering a chemical shift dispersion range of $^{14}$N of about 900 ppm\cite{Witanowski1993}, spectral resolutions are sufficient to resolve individual chemical shifts of the $^{14}$N spin system under pressure. A gradual shift of the signals towards lower ppm values was observed which could indicate changes in local environment of the $^{14}N$ nuclei. Details of our further analysis will be presented at a later stage.  

\subsection{Two dimensions}

While one-dimensional LG-NMR experiments prove to be a powerful tool to adequately resolve chemical shifts of homogeneous sample environments in DACs, signal detection in very heterogeneous systems or in systems with a many-fold of different chemical environments or phases remains difficult.

In order to deal with strongly overlapping spectra of different local chemical environments in one phase or coexisting phases, a deconvolution in a second or even  higher dimensions is a well established method in NMR spectroscopy \cite{Macomber1999}. In the case of Lee-Goldburg based homonuclear decoupling experiments, a second Fourier transform in the indirect LG-time domain correlates both laboratory and rotating frame free induction decays and is therefore an obvious choice for a second dimension.

The advantage of using such 2D high-resolution NMR spectra compared to their 1D analogue is that they allow for parallel investigation of static line shapes in the $F_1$ projection, indicating e.g. proton mobility or coordination changes in quadrupolar nuclei via changes in the charge symmetry, and detection of isotropic chemical, Knight or paramagnetic shifts in the $F_2$ projection.     
Three exemplary scenarios will be used in order to illustrate the usefulness of this multidimensional NMR technique in high pressure research.

\subsubsection{2D-LG-NMR on ferromagnetic $\delta-(Al_{0.3},Fe_{0.7})OOH$}

\begin{figure}
\includegraphics[width=0.45\textwidth]{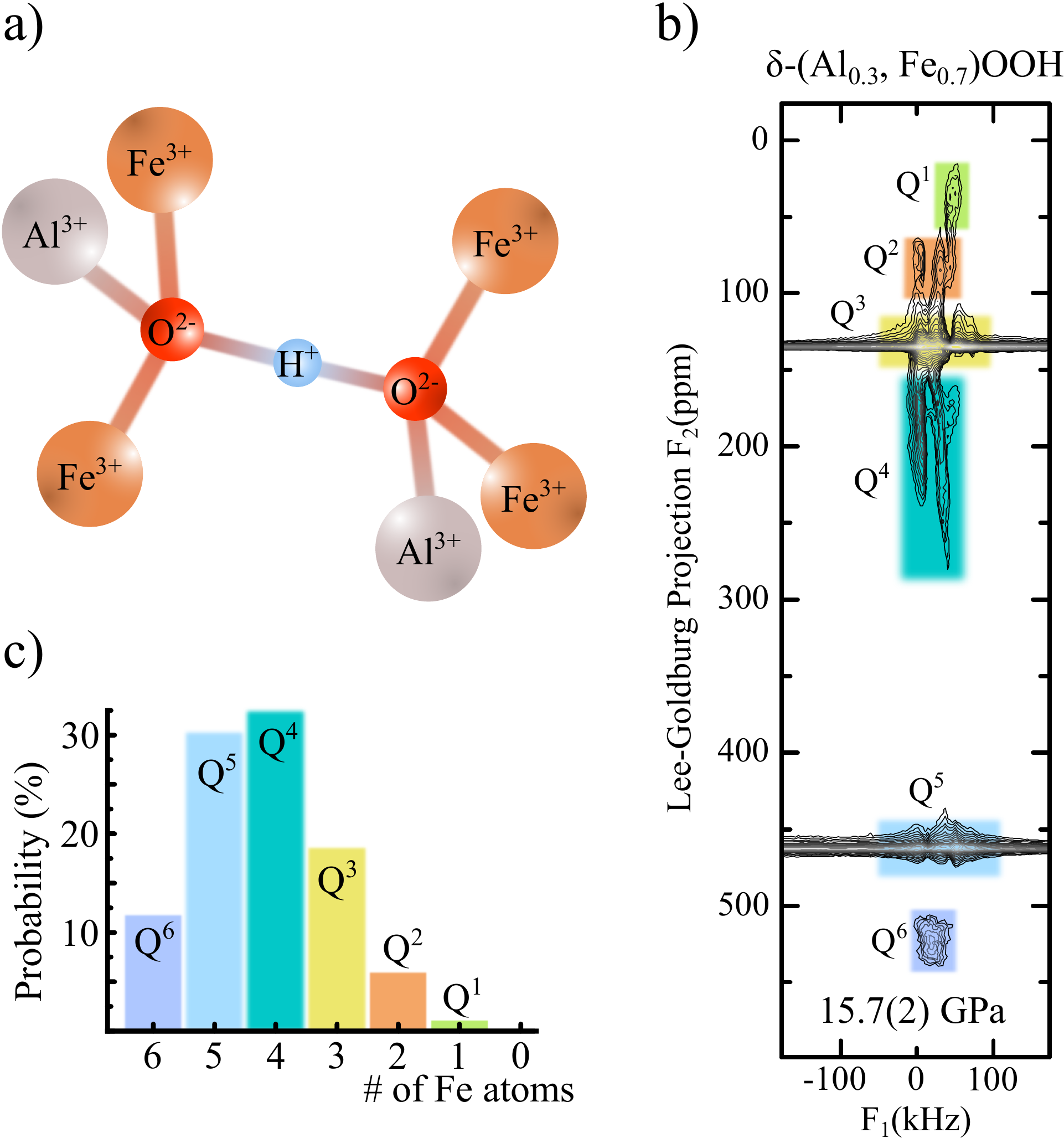}%
\caption{\textbf{High resolution 2D-LG-NMR on ferromagnetic $(Al_{0.3},Fe_{0.7})OOH$.} \textbf{a}) Local atomic structure of the hydrogen bond ensembles in $(Al_{0.3},Fe_{0.7})OOH$. As both $Al^{3+}$ and $Fe^{3+}$ cations are statistically distributed, several H-bond environments are likely to appear in the $^{1}$H-NMR spectra. \textbf{b}) 2D-$^{1}$H LG-NMR spectrum at 15.7 GPa. The hydrogen bond species Q$^{i}$, $i=1-6$ appear at different paramagnetic shift values in the indirect Lee-Goldburg projection dimension. Using the probability distribution of these species according to the stochiometry of the sample (\textbf{c)}), signal assignment via intensity ratios was possible.
\label{fig:oxyhydroxide}}
\end{figure}
Pure $\delta$-AlOOH undergoes a subtle sub to supergroup phase transition from the P2$_1$nm to the Pnnm phase at $\sim 10$ GPa \cite{Simonova2020,SF2018} and shows a hydrogen bond symmetrization at about 15 GPa \cite{SF2018, pillai2018,cortona2017}. This oxyhydroxide gained attention in the geophysics community due to its potential role as a carrier of hydrogen into the deeper regions of Earth's mantle \cite{sano2008,duan2018phase,su2021effect, Mao2020}.  Iron incorporation (replacing Al atoms with Fe) is expected under mantle conditions and introduces a spin transition from a $S=5/2$ high spin state to a  $S=1/2$ low spin state,between 37 and 48 GPa \cite{su2021effect} and may influence the structural phase transition, as well as, the hydrogen bond symmetrization. 

In Figure \ref{fig:oxyhydroxide}a), the local surrounding of the O-H-O bond is visualized with a hydrogen atom at its center. Each oxygen is bound to three metal atoms, being either Al$^{3+}$ or Fe$^{3+}$ cations, depending on the iron content. Assuming a stochastic distribution, six different hydrogen bond environments should be present. Figure \ref{fig:oxyhydroxide}c shows the theoretical probability distribution according to the stoichiometry of our $\delta$-(Al$_{0.3}$,Fe$_{0.7}$)OOH sample (Synthesis of the single crystal is explained in detail\citet{Simonova2020}). The hydrogen bond species $Q^i$ (i=1-6), denote the amount of ferric iron cations bound to the O-H-O subsystem.  

Figure \ref{fig:oxyhydroxide}b shows a high-resolution 2D-LG $^{1}$H-NMR spectrum of $\delta$-(Al$_{0.3}$,Fe$_{0.7}$)OOH at 15.7 GPa. Inspection of the area at $F_1 \simeq 20$ kHz revealed a number of additional, formerly unresolvable signals. The two-dimensional peak intensities of these additional signals were found to follow the probability distribution of the $Q^i$ species, allowing for signal assignment in the 2D spectrum.

Since the isotropic part of the paramagnetic shielding tensor is proportional to the total electron spin number $S$ and the local electron density of the paramagnetic centres at the NMR probe nulceus \cite{Pell2019}, it is reasonable to assume that the resonance signals shift to higher ppm values with the amount of ferric iron in the close surrounding, starting with Fe$^{3+}$ depleted H-bond environments ($Q^{0;1}$) akin to pure $\delta$-AlOOH to fully Fe$^{3+}$ occupied coordination shells similar to $\epsilon$-FeOOH. It is therefore possible to distinguish different local environments in the 2D-LG decomposition, enabling the investigation of the individual behaviour of the spin-subsystems as a function of the local iron content. A closer analysis of the high-pressure behaviour of these systems will be presented elsewhere.  
\\ 
\subsubsection{Spin transition in dense magnesium silicate phase D}
\begin{figure}[htb]
\includegraphics[width=0.4\textwidth]{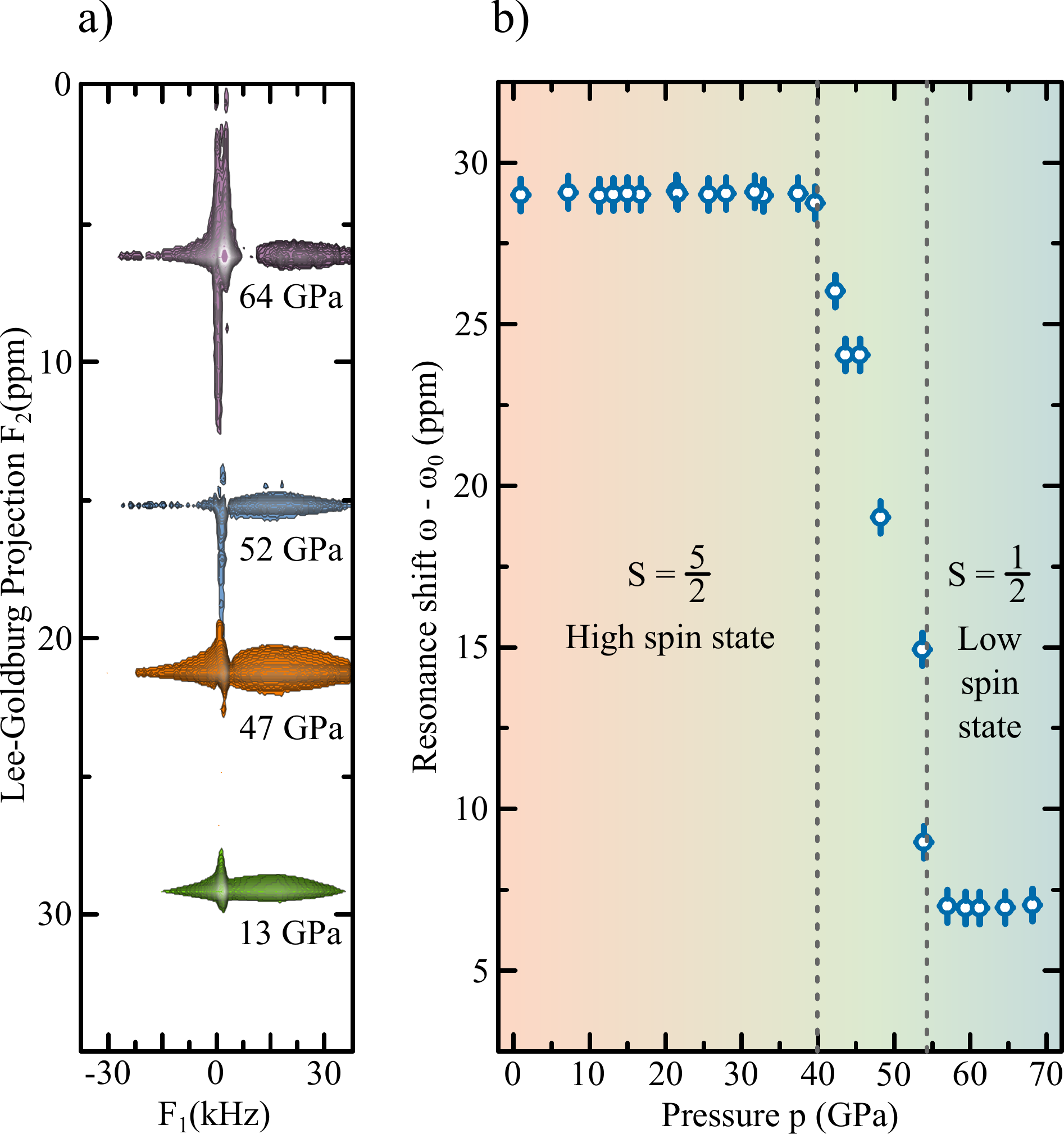}%
\caption{ \textbf{2D-$^{1}$H-LG-NMR spectra of dense magnesium silicate Phase D, $(Mg_{0.88},Fe_{0.12})\cdot(Si_{0.9},Al_{0.1})_2 O_6 H_2$.} \textbf{a)} The spectra have line-widths in the LG projection dimension of less than 1 ppm, allowing for the observation of the high spin to low spin transition of the ferric $Fe^{3+}$ ions resulting in a partial collapse of the paramagnetic shift interaction with the hydrogen nuclei. b) Resonance shift $\omega - \omega_0$ at pressures between 70 GPa and ambient conditions. Under the influence of strong paramagnetic interactions in the HS state below 40 GPa, the $^{1}H$-NMR signals of Phase D are shifted by 30 ppm downfield (higher ppm values). The electron spin crossover to a LS configuration leads to a pronounced reduction in the paramagnetic shift at higher pressures.} 
 \label{fig:phaseD}
 \end{figure} 
Two dimensional high-resolution signal detection in DACs can also be employed to observe otherwise undetectable electronic transitions. Figure \ref{fig:phaseD} shows 2D-LG $^{1}$H-NMR spectra of dense magnesium silicate phase D, (Mg$_{0.88}$,Fe$_{0.12}$)(Si$_{0.9}$,Al$_{0.1}$)$_2$O$_6$H$_2$)\cite{Yuan2017}. Due to the low concentration of iron in this sample, the HS$\rightarrow$ LS transition of ferric iron at $P\simeq 43$ GPa is not observable by diffraction methods via a volume collapse of the FeO$_6$ octahedra. Also, the presence of both divalent and trivalent iron in these samples complicates M\"ossbauer spectroscopy due to greatly overlapping signals, hindering a straightforward observation of the electron spin collapse.  
\\
On the contrary, $^{1}$H-NMR signals of the hydrogen atoms in phase D were found to be well resolved  in the indirect LG-frequency domain with line widths $\sim 1$ ppm. Furthermore, the presence of paramagnetic centres leads to a significant enhancement of spin-lattice relaxation times and thus allows for rapid data acquisition \cite{Pell2019}. The position of the resonances shift towards lower ppm values, from initially about 30 ppm at  up to 40 GPa to 6 ppm at higher pressures (Figure \ref{fig:phaseD}b) ). This transition was not observable using standard NMR excitation pulse sequences as the resonance line widths where found to be in excess of 200 ppm. Hydrogen signals stemming from paramagnetic interactions with high spin divalent iron atoms ($S=2$) were not observed within the chosen spectral range ($\approx 300$ ppm).

\subsubsection{Phase heterogeneity in metal hydrides}

One of the most intriguing discoveries in the past years has been the observation of near room temperature superconductivity in metal hydrides\cite{Ashcroft2004} based on first-principle considerations\cite{Chen2020}. First detected in H$_3$S at 204 K at $P \simeq 150$ GPa\cite{Drozdov2015}, it was rapidly realized that hydrogen incorporation into metallic parent lattices under high-$P$ high-$T$ conditions may be a route to a novel family of high-temperature superconductors \cite{Lv2020}. Recently, evidence for cooper pair condensation was found in clathrate-like superhydrides LaH$_{10}$  \cite{Drozdov2019, Somayazulu2018} as well as in carbonaceous sulfur hydrides\cite{Snider2020}.

However, sample synthesis and structure determination in DACs at the pressures ($\gtrsim 100$ GPa) necessary for the stabilization of these hydride structures are often challenging\cite{Struzhkin2020} and vastly limited to diffraction methods probing only the metal parent lattices or transport measurements \cite{Kong2019}. 

NMR spectroscopy could yield information about the electronic and dynamic properties of the hydrogen subsystems in these compounds. Our first experiments on iron \cite{Meier2019a} as well as copper hydrides \cite{Meier2020a} demonstrated that NMR can be used to detect hydrogen systems associated to metallic metal hydride phases synthesised at high-$P$ under laser heating conditions and that in combination with density-function-theory (DFT) based electronic structure calculations different coexisting hydrides can be distinguished from their electronic properties, if signals do not strongly overlap  \cite{Meier2020a} .

However, laser heated sample synthesis is prone to phase heterogeneity, where multiple metastable phases, which are similar in their electronic structure, can be synthesised along the steep thermal gradient of laser focus spots\cite{Kong2019}, which leads potentially to substantial signal overlap of static $^{1}$H-NMR spectra. 2D-LG-NMR  can be used in this vein to not only resolve signals from hydride phases which vary only slightly in their hydrogen content, but also to detect low abundances of such phases as the NMR signal intensity is in general proportional to the absolute abundance of each spin system or structure.
\begin{figure}[htb]
\includegraphics[width=0.45\textwidth]{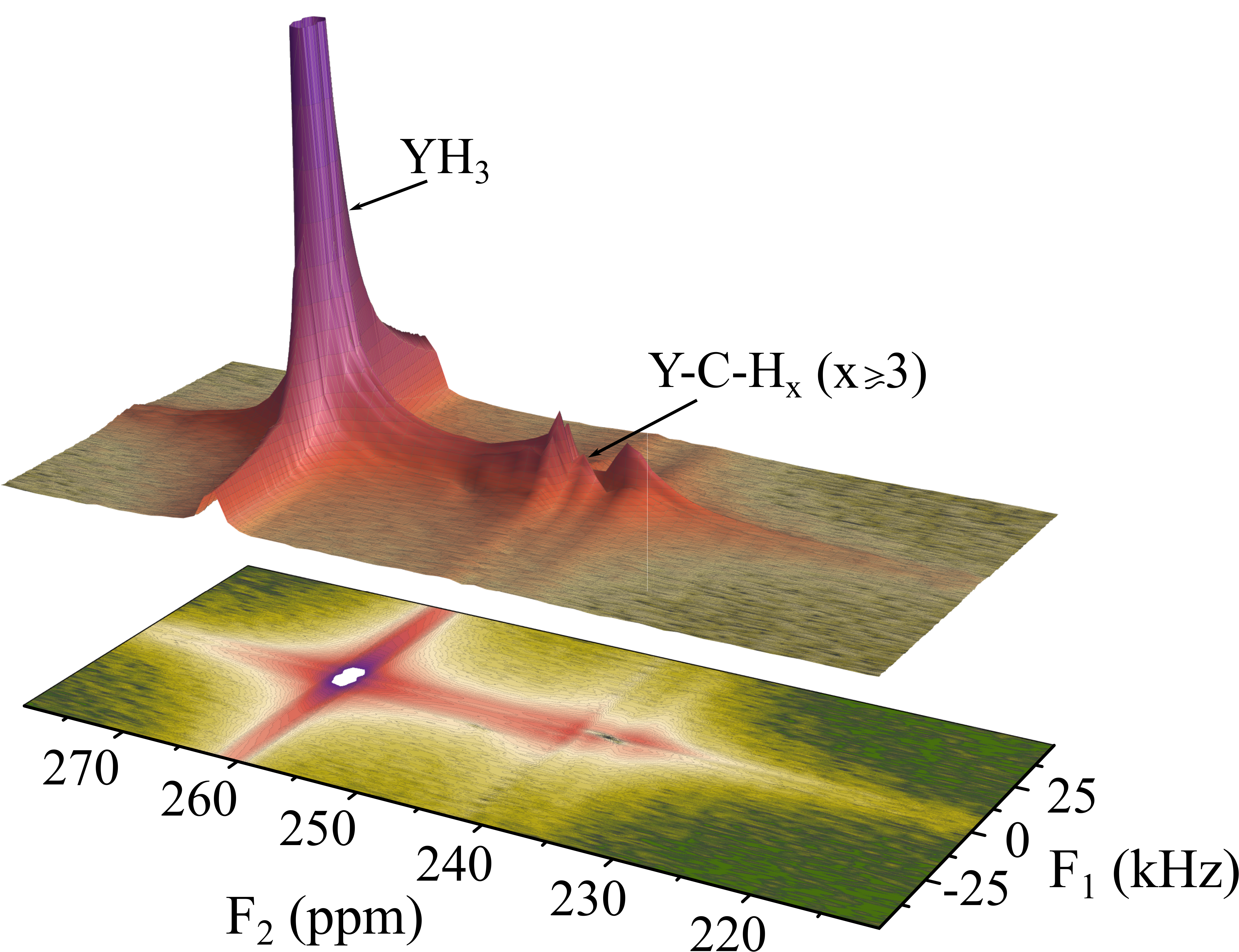}%
\caption{\textbf{High resolution $^{1}$H-NMR spectrum of yttrium hydrides synthesised at 45 GPa and 2500 K.} The spectrum of $YH_2$ appears at 0 ppm and is not shown. Despite synthesised $YH_3$, the formation of at least one other hydride of slighty higher hydrogen content than $YH_3$ or carbon incorporation could be observed. 
\label{fig:YH}}
\end{figure} 
\\
Figure \ref{fig:YH} shows a high-resolution LG spectrum of Yttrium hydride synthesised from a mixture of Yttrium powder and paraffin oil at 45 GPa under laser heating at up to 2500 K. Diffraction experiments suggest the predominant presence of two yttrium hydrides at these conditions, non-metallic YH$_2$ and metallic YH$_3$ \cite{Li2015,Liu2017a}. The former will be, in absence of paramagnetic centers, dominated by diamagnetic chemical shift interactions within $F_2 \le~20~ppm$ while the latter, being metallic, will be under the strong influence of Pauli hyperfine interaction leading to a pronounced Knight shift. 

We find four peaks in the 2-D spectrum: a peak at a shift of 0 ppm, one pronounced peak at 263 ppm and at least two additional weak signals at slightly lower shifts (Figure \ref{fig:YH}). We attribute the signal at 0 ppm to non-metallic YH$_2$. The additional signals may stem from metallic Yttrium hydride phases and consequently the spin lattice relaxation times of these spin systems should follow the Korringa relation for Fermi-contact driven electron-nuclear hyperfine interaction via:
\begin{equation}
T_1=\frac{\hbar}{4\pi k_{\rm B}}\cdot \left( \frac{\gamma_{ \rm e}}{\gamma_{\rm n}}\right)^2 \cdot \frac{1}{K_{\rm s}^2 T}, 
\label{korringa}
\end{equation}
where $\gamma_{\rm e}$ and $\gamma_{\rm n}$ are the gyromagnetic ratios of the electron and the hydrogen nuclei respectively, $T$ the sample temperature and $K_{\rm s}$ the Fermi-contact term of the Knight shift. 

Using equation \eqref{korringa}, spin lattice relaxation times $T_1$ in the order of 10 ms could be anticipated.

Using saturation recovery experiments, we found $T_1$ values of 8 to 12 ms in excellent agreement with the theoretical Korringa relation, evidencing the metallic character of the observed signals (Figure \ref{fig:YH}). Therefore, we attribute the signal at 263 ppm to YH$_3$. The presence of  weaker signals in addition to the expected signals of YH$_2$ and YH$_3$ in the high-resolution Lee-Goldburg spectrum suggests the formation of either YH$_{\rm x}$ phases with slightly higher hydrogenation \cite{Meier2020a} or possible metallic ternary compounds formed by incorporation of carbon stemming from the paraffin hydrogen reservoir or from the diamond anvils. Further investigation of the reaction dynamics as well as supporting DFT-based computations will hopefully help to identify these additional signals and will be presented at a later stage. 

Nevertheless, this preliminary analysis of the data shows, that it is possible to deconvolute NMR signals of several metal hydride phases simultaneously synthesized in one DAC. Therefore, limitations arising from non phase pure synthesis in other experimental techniques do not influence LG-NMR experiments in the same way, leading to the opportunity to simultaneously investigate several coexisting metallic and possible superconducting phases at the same $P$ and $T$ conditions.
\section{Conclusion}
In this work, we presented the first high-resolution high-pressure Nuclear Magnetic Resonance experiments in diamond anvil cells using the well established method of a rotating radio frequency magnetic field aligned in the magic angle relative to the external magnetic field. We show that this method for resonance line narrowing is particularly suitable for our used Lenz lens based high-pressure approach to NMR spectroscopy, leading to an improvement of the spectral resolution to 1 ppm and below. Several contemporary applications from the fields of condensed matter research, geo- and solid-state physics\cite{Mao2021} as well as high pressure chemistry\cite{Yoo2020} demonstrate the vast applicability of this method. 

Furthermore, additional developments, not presented in this work, e.g.
\begin{description}
\item[(i)] further improvements of the sensitivity for small sample fractions in the 2D spectra
\item[(ii)] the combination of high-$P$ NMR with structure search calculations as suggested by Monserrat \textit{et al.}\cite{Monserrat2019}
\item[(iii)] phase sensitive measurements of superconducting properties.
\end{description}
will hopefully lead to a significant broadening of the usage of NMR spectroscopy in high-pressure research and establish high-pressure NMR crystallography as an additional experimental tool complementing diffraction methods, in particular with respect to low Z-materials\cite{Ji2020} and DFT-based calculations\cite{Semenok2018}.

\section{Acknowledgements}
We thank Nobuyoshi Miyajima for help with the FIB milling. We thank the German Research Foundation (Deutsche Forschungsgemeinschaft, DFG, Project Nos. DU 954/11-1, DU 393/13-1, DU 393/9-2 and ME 5206/3-1) and the Federal Ministry of Education and Research, Germany (BMBF, Grant No. 05K19WC1) for financial support. FT thanks the Swedish Research Council (VR) (Grant No. 2019-05600). D.L. thanks the Alexander von Humboldt Foundation for financial support. N.D. thanks the Swedish Government Strategic Research Area in Materials Science on Functional Materials at Linköping University (Faculty Grant SFO-Mat-LiU No. 2009 00971).
\section{Funding}
The authors thank the German Research Foundation (Deutsche Forschungsgemeinschaft, DFG, Project Nos. DU 954/11-1, DU 393/13-1, DU 393/9-2 and ME 5206/3-1) and the Federal Ministry of Education and Research, Germany (BMBF, Grant No. 05K19WC1) for financial support. T.M. thanks the Center for High Pressure Sience and Technology Advanced Research for financial support. 
F.T. thanks the Swedish Research Council (VR) (Grant No. 2019-05600). D.L. thanks the Alexander von Humboldt Foundation for financial support. N.D. thanks the Swedish Government Strategic Research Area in Materials Science on Functional Materials at Linköping University (Faculty Grant SFO-Mat-LiU No. 2009 00971).
\\
%

\end{document}